# Dependence of the Energy Spectrum of UHE Cosmic Rays on the Latitude of an Extensive Air Shower Array

A.V. Glushkov

*Institute of Cosmophysical Research and Aeronomy, Yakut Research Center, Siberian Branch, Russian Academy of Sciences, pr. Lenina 31, Yakutsk, 677891 Russia*
*e-mail: a.v.glushkov@ikfia.ysn.ru*

Several energy spectra of cosmic rays with energies $E_0 \geq 10^{17}$ eV measured at the Yakutsk EAS, AGASA, Haverah Park, HiRes, Auger, and SUGAR arrays are considered. It is shown that the fairly good mutual agreement of the spectrum shapes can be achieved if the energy of each spectrum is multiplied by a factor $K$ specific for each spectrum. These factors exhibit a pronounced dependence on the latitude of the above-mentioned arrays.

## 1. INTRODUCTION

The energy spectrum of ultrahigh-energy (UHE) cosmic rays ($E_0 \geq 10^{17}$ eV) is a key link in the chain of difficult problems associated with the insight into the origin of the primary particles of these energies. Giant air showers (GASs) with ultimate energies $E_0 \geq 10^{19}$ eV have attracted great interest for almost four decades since the first such events have been detected by the largest world arrays Volcano Ranch (USA) [1], Haverah Park (England) [2], SUGAR (Sydney University Giant Air Shower Recorder) [3], and Yakutsk array [4]. In many respects, this interest is stimulated by the drastic change in the spectrum shape observed at all arrays at these energies, as well as by the strong differences between the intensities of different spectra. This is clearly seen in Fig. 1a showing the statistically significant results obtained by 1987. The open circles represent the SUGAR differential spectrum published in its final form in [5] after this array has been closed; the closed circles and the open and closed triangles show the Yakutsk spectrum [6]; and the crosses, the Haverah Park spectrum [7]. The sloping straight line drawn via the characteristic dips of the spectra corresponds to an intensity change by a factor of $3\Delta\log E_0$ upon an energy change by a factor of $\Delta\log E_0$. The SUGAR and Haverah Park spectra comprise events with $E_0 \geq 10^{20}$ eV, which contradict the theoretically expected Greisen–Zatsepin–Kuzmin suppression of the spectrum at $E_0 \geq 6 \times 10^{19}$ eV (the GZK cutoff) due to the intense interaction of GASs with the cosmic microwave background [8, 9]. The Yakutsk spectrum exhibits a distinct dip for $E_0 = 8 \times 10^{18}$ eV and a bump for $E_0 \approx 5 \times 10^{19}$ eV and does not contradict the GZK cutoff at the higher energies.

The problem of the GZK cutoff stimulated the development of a number of new, more sensitive arrays such as AGASA (Akeno Giant Air Shower Array, Japan) and HiRes Fly's Eye (USA). The final solution of the problem of GAS origin was expected to be finally solved at these arrays, but this expectation has not been realized. A new intrigue arose after the supergiant Auger array (3000 km² in area) has been recently put into operation in Argentine. A number of sensational reports based on its results were presented at the 30th International Cosmic Ray Conference in Mexico in the summer of 2007 (see reviews [10, 11]).

In this paper, the problem of the UHE cosmic-ray spectrum is analyzed within the framework of its historical context, without discriminating even the results which seem to be out-of-date or not very reliable. This makes the picture of this problem more comprehensive and opens up certain new ways for its solution.

## 2. EXPERIMENTAL RESULTS AND DISCUSSION

As seen in Fig. 1a, all of the three spectra have fairly similar shapes. Recall that the primary-particle energies at the SUGAR array were found based on the total number of vertical muons with the threshold energy $E_\mu = 0.75 \sec\vartheta$ GeV, while the density $\rho_{600}$ of charged particles measured by ground-based Cherenkov water tanks with a thickness of 1.2 m at a distance of 600 m from the EAS axis was used for this purpose at the Haverah Park array. At the Yakutsk array, $E_0$ was deduced from $S_{600}$, the density of charged particles detected by ground-based scintillation detectors with a thickness of 5 cm and an area of 2 m² at a distance of 600 m from the EAS axis.

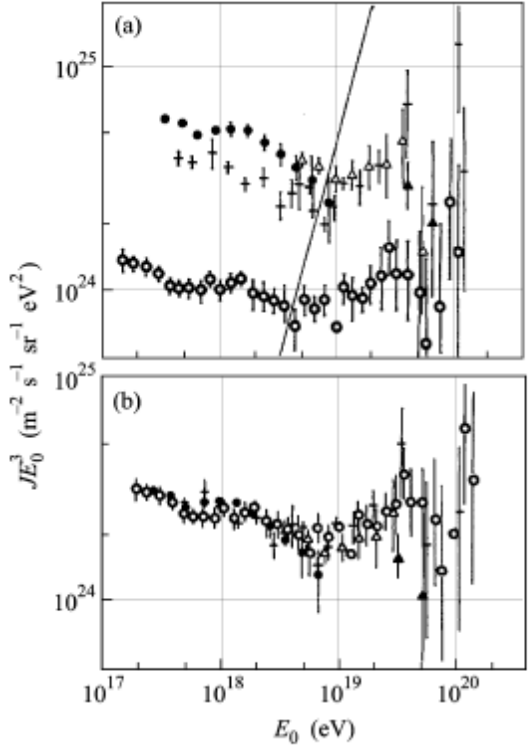

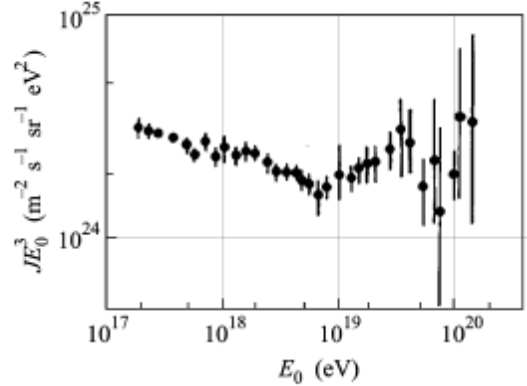

**Fig. 1.** (a) Differential energy spectra measured (○, [5]) at the SUGAR array for showers with zenith angles $\vartheta \leq 60°$, (+, [7]) at the Haverah Park arrays, (●) at the small (500 m) and (△) large (1000 m) masters for showers with $\vartheta \leq 45°$ detected inside the Yakutsk array, and (▲, [6]) for the showers with $\vartheta \leq 60°$ detected at the Yakutsk array with an extended area of 30 km². (b) The same spectra upon scaling their energies by the factors $K = 1.29, 0.9,$ and $0.81$, respectively.

**Fig. 2.** Differential energy spectrum upon averaging over individual data points in Fig. 1b.

Similar to [12], one can assume that the observed difference in the spectrum intensities is related to some methodological features of various experiments; e.g., to uncertainties in the determination of $E_0$. Let us try to make the spectra in Fig. 1a maximally close to each other by parallel translations along the sloping straight line to a certain position. Let this common position be chosen such that the power-law fits $J(E_0) \sim E_0^{-\gamma}$ in the energy ranges $(1-8) \times 10^{18}$ and $(8-50) \times 10^{18}$ eV yield the intensities $JE_0^3 = 2 \times 10^{24}$ m$^{-2}$ s$^{-1}$ sr$^{-1}$ eV² for $E_0 = 3.15 \times 10^{18}$ and $1.4 \times 10^{19}$ eV, respectively, while the spectrum exponents in these energy ranges are $\gamma = 3.2$ and $2.68$, respectively. This requires the Yakutsk, Haverah Park, and SUGAR energies to be scaled by the factors $K = 0.81 \pm 0.05, 0.9 \pm 0.07,$ and $1.29 \pm 0.1$, respectively. The values of $K$ and the corresponding errors are determined using the $\chi^2$ test. The result of such a procedure is shown in Fig. 1b. The shapes of all of the three spectra are in fairly good agreement in the entire UHE energy range. By averaging individual points, we arrive at the spectrum shown in Fig. 2. It exhibits a clearly pronounced dip for $E_0 \approx 8 \times 10^{18}$ eV and a bump for $E_0 \approx 4 \times 10^{19}$ eV. A few giant air showers with $E_0 > 10^{20}$ eV impede solving the problem of the existence of the GZK cutoff at present.

This was the status of the ultrahigh-energy cosmic-ray spectrum when the new AGASA [13], HiRes [14], and Auger [10] arrays were put into operation. The spectra obtained at these arrays are shown in Fig. 3a together with the spectra from Fig. 1a supplemented by the most recent Yakutsk data [15]. Note that the AGASA scintillation detectors and technique employed for estimating $E_0$ from $S_{600}$ were similar to those used at the Yakutsk array. This probably explains the fairly close intensities and shapes of the spectra at these two arrays. An exception is constituted by 11 GASs with $E_0 \geq 10^{20}$ eV detected at the AGASA array and only one such event detected at the Yakutsk array. At the HiRes and Auger arrays, the energy of the primary particles was determined using the atmospheric fluorescence accompanying the EAS development. However, the energies of the overwhelming majority of GASs detected at the Auger array were still determined using $\rho_{1000}$, the charged-particle density at a distance of 1000 m from the EAS axis, measured by Cherenkov water tanks with a thickness of 1.2 m.

It is seen that the new spectra complicate the general picture. First, the strong difference between the HiRes and Auger spectra intensities seems quite strange, because the energy of the primary particles was estimated in both cases using the same technique based on EAS fluorescence. Second, the difference between the Haverah Park and Auger spectra is even more surprising, since water Cherenkov tanks of similar types were used as the main detectors at these arrays. Watson, the leader of these arrays, comments nothing in this respect. Third, the Auger and SUGAR spectra turn out to be surprisingly similar. The very low intensity dis-credited the latter in due time, and, unfortunately, it has been forgotten for quite some time. However, the SUGAR spectrum has received a worthwhile successor. Such a wide variety of spectra stimulates attempts to find a general reason for their disagreement.

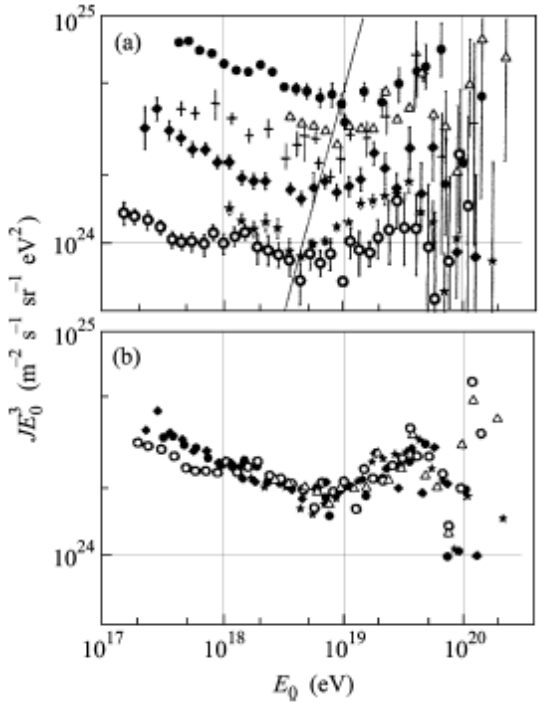

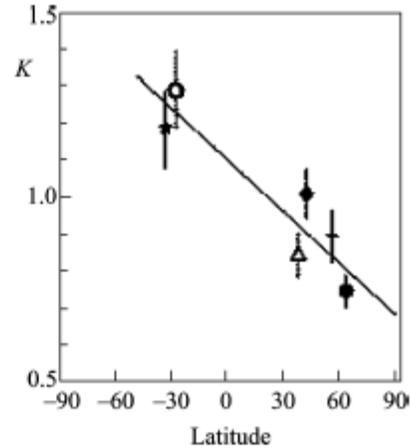

**Fig. 4.** Normalization factors versus latitudes of the arrays (the notation is the same as in Fig. 3).

**Fig. 3.** (a) Differential energy spectra measured at (●, [15]) the Yakutsk array for zenith angles $\vartheta \leq 60°$, (△, [13]) the AGASA array for $\vartheta \leq 60°$, (+, [7]) the Haverah Park array for $\vartheta \leq 60°$, (◆, [14]) the HiRes array, (★, [10]) the Auger array for $\vartheta \leq 70°$, and (○, [5]) the SUGAR array for $\vartheta \leq 60°$. (b) The same spectra upon scaling their energies by the factors $K = 0.75, 0.85, 0.9, 1.02, 1.19,$ and $1.29$, respectively.

It is easily seen that all of the spectra in Fig. 3a have fairly similar shapes. This becomes obvious if the energies of the spectra measured at the Yakutsk, AGASA, HiRes, and Auger arrays are scaled by the factors $K = 0.75 \pm 0.04, 0.85 \pm 0.06, 1.02 \pm 0.07,$ and $1.19 \pm 0.11,$ respectively. The results of such a scaling are shown in Fig. 3b. For clarity, the errors are not plotted in this figure. Such a relative normalization of spectra was proposed by Berezinsky et al. [12] who believe that the absolute flux of cosmic rays itself is isotropic, and the problem of the disagreement of various spectra is essentially related to the errors of energy estimation employed at various arrays. This seems reasonable to some extent. However, we point to an interesting fact following from Fig. 3a and illustrated in Fig. 4. The latter figure demonstrates a certain dependence of the normalization factors $K$ on the latitudes of the arrays considered. The further south the array location, the lower the corresponding spectrum. It is difficult to treat such a relation as random, i.e., caused entirely by methodological errors of the determination of $E_0$. Most probably, it has a different origin.

## 3. CONCLUSIONS

Six UHE cosmic-ray spectra measured at different EAS arrays worldwide at different times are considered. These spectra attracting permanent interest have fairly high statistical significance. They are based on the detection of various EAS components and obtained using different data-processing techniques. These spectra turn out to be different (Fig. 3a). The HiRes and Auger results have attracted the greatest interest in recent years; moreover, some specialists prefer to focus discussions only on these results. In our opinion, any discrimination can hardly facilitate progress on this issue. Deeper insight requires the application of a complex approach to understand the entire data set. The spectrum normalization method [12] provides a good agreement between the shapes of all spectra (see Fig. 3b). However, differences in the intensities of the spectra are most probably caused not by energy-calibration errors at different arrays, but by the fact that these arrays cover different sky regions. Using the Yakutsk data, we already showed in [16, 17] that energy spectra from different sky regions are notably different. All of this requires additional careful analysis, which we will continue.